\documentclass[preprint2]{emulateapj}
\usepackage{times}

\newcommand{\EQ}{\begin{equation}}
\newcommand{\EN}{\end{equation}}
\newcommand{\EQA}{\begin{eqnarray}}
\newcommand{\ENA}{\end{eqnarray}}
\newcommand{\eq}[1]{(\ref{#1})}

\newcommand{\Eq}[1]{equation~(\ref{#1})}

\newcommand{\meanemf}{\overline{\mbox{\boldmath ${\cal E}$}}{}}{}
\newcommand{\meanFF}{\overline{\mbox{\boldmath ${\cal F}$}}{}}{}
\newcommand{\meanemfs}{\overline{\cal E} {}}

{}
{}
{}
{}
{}
{}
{}
{}
\newcommand{\meanBB}{\overline{\mbox{\boldmath $B$}}{}}{}
\newcommand{\meanJJ}{\overline{\mbox{\boldmath $J$}}{}}{}
\newcommand{\meanUU}{\overline{\mbox{\boldmath $U$}}{}}{}

\newcommand{\meanB}{\overline{B}}

\newcommand{\meanU}{\overline{U}}

\newcommand{\meanF}{\overline{\cal F}}
\newcommand{\rrr}{\mbox{\boldmath $r$} {}}
\newcommand{\yy}{\mbox{\boldmath $y$} {}}

\newcommand{\RR}{\mbox{\boldmath $R$} {}}
\newcommand{\kk}{\mbox{\boldmath $k$} {}}
\newcommand{\KK}{\mbox{\boldmath $K$} {}}
\newcommand{\uu}{\mbox{\boldmath $u$} {}}
\newcommand{\UU}{\mbox{\boldmath $U$} {}}
\newcommand{\xx}{\mbox{\boldmath $x$} {}}
\newcommand{\bb}{\mbox{\boldmath $b$} {}}
\newcommand{\BB}{\mbox{\boldmath $B$} {}}
\newcommand{\jj}{\mbox{\boldmath $j$} {}}
\newcommand{\JJ}{\mbox{\boldmath $J$} {}}
\newcommand{\SSS}{\mbox{\boldmath $S$} {}}
\newcommand{\AAA}{\mbox{\boldmath $A$} {}}
\newcommand{\aaaa}{\mbox{\boldmath $a$} {}}
\newcommand{\ee}{\mbox{\boldmath $e$} {}}

\newcommand{\nab}{\mbox{\boldmath $\nabla$} {}}

\newcommand{\ii}{{\rm i}}

\newcommand{\dd}{{\rm d} {}}

\def\half{{\textstyle{1\over2}}}

\newcommand{\yjgr}[3]{ #1, {J.\ Geophys.\ Res.,} {#2}, #3}

\newcommand{\yapj}[3]{ #1, {ApJ,} {#2}, #3}

\newcommand{\yan}[3]{ #1, {Astron.\ Nachr.,} {#2}, #3}

\newcommand{\yana}[3]{ #1, {A\&A,} {#2}, #3}
\newcommand{\yanas}[3]{ #1, {A\&AS,} {#2}, #3}

\newcommand{\yjfm}[3]{ #1, {J.\ Fluid Mech.,} {#2}, #3}

\newcommand{\yprl}[3]{ #1, {Phys.\ Rev.\ Lett.,} {#2}, #3}

\newcommand{\ymn}[3]{ #1, {MNRAS,} {#2}, #3}

\newcommand{\ysph}[3]{ #1, {Solar Phys.,} {#2}, #3}

\newcommand{\yjour}[4]{ #1, {#2}, {#3}, #4}

\newcommand{\ybook}[3]{ #1, {#2} (#3)}

\begin{document}

\title{Magnetic Helicity Density and Its Flux in Weakly Inhomogeneous Turbulence}
\author{Kandaswamy Subramanian\altaffilmark{1}
and Axel Brandenburg\altaffilmark{2}}
\altaffiltext{1}{Inter-University Centre for Astronomy and Astrophysics,
Post bag 4, Ganeshkhind, Pune 411 007, India;
\url{kandu@iucaa.ernet.in}}
\altaffiltext{2}{Nordic Institute for Theoretical Physics (NORDITA),
Blegdamsvej 17, DK-2100 Copenhagen \O, Denmark;
\url{brandenb@nordita.dk}}

\begin{abstract}
A gauge invariant and hence physically meaningful
definition of magnetic helicity density for
random fields is proposed, using the Gauss linking formula,
as the density of correlated field line linkages.
This definition is applied to the random small scale field
in weakly inhomogeneous turbulence, whose correlation length is small
compared with the scale on which the turbulence varies.
For inhomogeneous systems, with or without boundaries,
our technique then allows one to study
the local magnetic helicity density evolution
in a gauge independent fashion, which was not possible earlier.
This evolution equation is
governed by local sources (owing to the mean field) and by the divergence
of a magnetic helicity flux density.
The role of magnetic helicity fluxes in alleviating catastrophic quenching
of mean field dynamos is discussed.
\end{abstract}


%\maketitle

\section{Introduction}

Large scale magnetic fields produced by dynamo action tend to have some
degree of magnetic helicity.
A simple example is the interlocking of poloidal and toroidal fields in
one hemisphere of a star, seen from stellar dynamo models.
However, in the case of the Sun there is explicit evidence of
magnetic helicity being present in or coming from active regions
(Seehafer 1990; Pevtsov et al.\ 1995; Bao et al.\ 1999;
Berger \& Ruzmaikin 2000),
coronal mass ejections (D\'emoulin et al.\ 2002),
and the solar wind (Matthaeus et al.\ 1982; Lynch et al.\ 2005).
While the investigation of magnetic helicity in the Sun and in the
solar wind is interesting in its own right, there is now also a direct
connection with dynamo theory with the realization that large
scale dynamos must transport and
shed small scale magnetic helicity in order to
operate on a dynamical time scale rather than the much longer resistive
time scale (see the review of Brandenburg \& Subramanian 2005a for references).
A major difficulty with this picture is the absence of a meaningful
definition for magnetic helicity {\it density},
even for small scale fields.
Magnetic helicity is a volume integral, 
usually defined as $H_{\rm M}=\int\AAA\cdot\BB\,\dd V$,
where $\AAA$ is the vector potential and $\BB=\nab\times\AAA$ is 
the magnetic field.
However, under a gauge transformation $\AAA' = \AAA +\nab\Lambda$, which
leaves $\BB$ invariant, one has $H'=H+\int\Lambda\BB\cdot\dd\SSS$.
So $H$ is only gauge invariant if the $\BB$ field
has no component normal to the boundary or if it
vanishes sufficiently rapidly at the boundary of the integration volume.
In most practical contexts, like the Sun or galaxies,
however, the field does not vanish on the boundaries.
A possible remedy might be to consider instead
the gauge-invariant {\it relative} magnetic helicity,
defined by subtracting the helicity of a reference vacuum field
in the same gauge
(Berger \& Field 1984, Finn \& Antonsen 1985).
But the flux of relative helicity is cumbersome
to work with for arbitrarily shaped boundaries.
Also the concept of a density of relative helicity is
not meaningful, since it is defined only as a volume integral.
Indeed, there is simply no way that the quantity $\AAA\cdot\BB$ itself
can be gauge-invariant.

On an earlier occasion, Subramanian \& Brandenburg (2004, hereafter SB04)
considered the
evolution of the current helicity density, $H_{\rm C}=\JJ\cdot\BB$, where
$\JJ=\nab\times\BB/\mu_0$ is the current density and $\mu_0$ is the
vacuum permeability (we set $\mu_0=1$ in what follows).
Note that $H_{\rm C}$, as well as its flux, are locally well defined,
explicitly gauge invariant and observationally measurable. 
Furthermore, from a closure model, Pouquet et al.\ (1976) show that
the $\alpha$ effect needed for large-scale dynamos 
has a nonlinear addition due to the small scale contribution to $H_{\rm C}$.
The build up of this small-scale current helicity then goes to cancel the
kinetic part of the $\alpha$ effect and causes catastrophic 
quenching of the dynamo, unless
one can have a helicity flux out of the system. This formed the motivation
for the work of SB04. The major disadvantage in working with $H_{\rm C}$ 
however
is that one loses the conceptually simple form of the magnetic helicity 
conservation law. We propose here instead an alternative
means to define magnetic helicity density for the random
small scale field, using the more basic Gauss linking
formula for helicity, which can be directly applied to
discuss magnetic helicity density and its flux even in
inhomogeneous systems with boundaries.
The technique applied in calculating the magnetic helicity evolution
is however very similar to that employed in SB04.

In the following we define random small scale quantities
as departures from the corresponding
mean field quantity, e.g., $\bb=\BB-\meanBB$ for the magnetic field,
$\jj=\JJ-\meanJJ$ for the current density,
and $\uu=\UU-\meanUU$ for the velocity.
Throughout this Letter we adopt ensemble averages that, in practice,
are commonly approximated as spatial averages over one or two coordinate
directions (see, e.g., Brandenburg \& Subramanian 2005a).
However, the approach developed below applies also to the case without
a mean magnetic field.
Therefore, specific applications to the mean field dynamo (MFD) will be
postponed until the end of this Letter.

\section{Magnetic helicity density}

Given the random small scale
magnetic field $\bb(\xx,t)$ one can also define the magnetic 
helicity directly in terms of the field, as the linkage of its flux,
using Gauss's linking formula (Berger \& Field 1984, Moffatt 1969)
\EQ 
h_{\rm G} = \frac{1}{4\pi} \int\int \bb(\xx)\cdot\left[\bb(\yy) \times 
\frac{\xx -\yy}{\vert\xx -\yy\vert^3} \right]
\dd^3x \ \dd^3y,
\label{gauss}
\EN
where both integrations extend over the full volume.
Suppose we define an auxiliary field
\EQ
\aaaa(\xx) 
= \frac{1}{4\pi}\int \bb(\yy) \times \frac{\xx -\yy}{{\vert\xx -\yy\vert}^3} \ \dd^3y,
\EN
then this field satisfies $\nab \times \aaaa = \bb$, and  
$\nab\cdot\aaaa=0$, and one can write $h_{\rm G} = \int \aaaa\cdot\bb \ \dd^3x$.
This is the origin of the textbook definition of magnetic helicity
in what is known as
the Coulomb gauge for the vector potential.
Provided the field is closed over the integration volume,
this definition can be applied in any other gauge.
Note however that for an open system with boundaries, 
it is {\it not} useful to go to the definition involving the 
vector potential, which is now of course gauge dependent.
We therefore take the point of view here that the magnetic helicity
density $h_{\rm G}$ 
defined by \Eq{gauss} is the more basic definition of the topological
property determining the links associated with magnetic fields,
and not the definition in terms of the vector potential.
We will see below that this then also allows us
to define naturally a gauge invariant magnetic helicity density for random
fields, as long as the correlation scale of the field is much
smaller than the size of the system, as the density of correlated 
links of the field.

For this consider $\bb$ to be a random field with a correlation
function $\overline{b_i(\xx,t)b_j(\yy,t)} = M_{ij}(\rrr,\RR)$.
Here we have defined the difference $\rrr = \xx - \yy$ and the mean
$\RR = (\xx + \yy)/2$, keeping in mind that,
for weakly inhomogeneous turbulence,
the two-point correlation
$M_{ij}(\rrr,\RR)$ and in fact all two-point correlations below,
vary rapidly with $\rrr$ but vary slowly with $\RR$.
Taking the ensemble average of $h_{\rm G}$, we have
\EQ
\overline{h}_{\rm G} = \frac{1}{4\pi} \int \dd^3R \int \dd^3r \;
\epsilon_{ijk} M_{ij}(\rrr,\RR)\,\frac{r_k}{r^3}.
\EN
Next, we suppose that the correlation scale $l$ of the random small scale field
$\bb$ is much smaller than the system scale $R_{\rm S}$, i.e.\ we
suppose that there exists an intermediate
scale $L$ such that $l \ll L \ll R_{\rm S}$ with
$M_{ij}(\rrr,\RR) \to 0$ as $\vert\rrr\vert \to L \gg l$.
Then one can do the $\rrr$ integral even by restricting oneself to the
intermediate scale $L$ and still capture all the
dominant contributions to the integral. This then motivates us to
define the magnetic helicity density $h$ of the random small scale field  
as $\overline{h}_{\rm G} = \int \dd^3R \ h(\RR)$. Here,
\EQ
h(\RR) = \frac{1}{4\pi} \int_{L^3} \dd^3r
\ \epsilon_{ijk} M_{ij}(\rrr,\RR)\,\frac{r_k}{r^3},
\label{heldef}
\EN
where we can formally let $L \to \infty$. The above expression
for $h(\RR)$ in \Eq{heldef} is our proposal
for the helicity density of the random small scale field $\bb$.
Evidently, $h(\RR)$ is explicitly gauge invariant.
A qualitative description would be to say that the magnetic helicity density
of a random small scale field is the density of correlated links
of the field.  We can now derive the evolution equation for $h(\RR)$
and also meaningfully (in a gauge invariant manner) talk about its flux.
Note that this has not been possible before, although many papers
(e.g., Blackman \& Field 2000; Kleeorin et al.\ 2000; Vishniac \& Cho 2001)
have appealed to
the notion of a magnetic helicity flux density in some qualitative fashion.
We will see that the magnetic helicity evolution equation that we derive
reproduces the known evolution equation
for homogeneous turbulence and generalizes it to the
inhomogeneous case by introducing possible fluxes of helicity.

\section{Magnetic helicity density evolution}

It is much simpler to work out the evolution equation for
$h(\RR)$ by first going to Fourier space, using 
the two-scale approach of Roberts \& Soward (1975), where
all two-point correlations are assumed to vary rapidly
with $\rrr$ and slowly with $\RR$.
Consider the equal time,
ensemble average of the product $\overline{f(\xx_1)g(\xx_2)}$.
The common dependence of $f$ and $g$ on $t$ is assumed
and will not explicitly be stated.
Let  $\hat{f}(\kk_1)$ and $\hat{g}(\kk_2)$ be the Fourier transforms
of $f$ and $g$, respectively.
We can express this correlation as
$\overline{f(\xx_1)g(\xx_2)}
= \int \Phi(\hat{f},\hat{g},\kk,\RR) \ {\rm e}^{i\kk\cdot\rrr} \
\dd^3k$, with 
\EQ
\Phi(\hat{f},\hat{g},\kk,\RR)
= \int \overline{\hat{f}(\kk + \half\KK) \hat{g}(-\kk +\half\KK)}
\ {\rm e}^{i\KK\cdot\RR} \,\dd^3K .
\label{phi_def}
\EN
Here, $\kk = \half(\kk_1 - \kk_2)$ and $\KK = \kk_1 + \kk_2$.
We define in Fourier space the correlation and cross correlation tensors
of the $\uu$ and $\bb$ fields;
$v_{ij}(\kk,\RR)=\Phi(\hat{u}_i,\hat{u}_j,\kk,\RR)$,
$m_{ij}(\kk,\RR)=\Phi(\hat{b}_i,\hat{b}_j,\kk,\RR)$, and
$\chi_{jk}(\kk,\RR)=\Phi(\hat{u}_j,\hat{b}_k,\kk,\RR)$.
In MFD theory, the turbulent electromotive force (EMF) is given by
$\meanemf=\overline{\uu\times\bb}$, whose components are
$\meanemfs_i(\RR) = \epsilon_{ijk} \int \chi_{jk}(\kk,\RR) \,\dd^3k$.
Furthermore, in Fourier space we have for the magnetic helicity density,
\EQA
h(\RR) = \int \int \epsilon_{ijk} \ 
 \overline{\hat{b}_i(\kk + \half\KK) \hat{b}_j(-\kk +\half\KK)}
\nonumber \\
\times (\ii k_k/k^2) {\rm e}^{i\KK\cdot\RR}
\ \dd^3k \ \dd^3K  .
\label{helfour}
\ENA
We should remark, that for an inhomogeneous system,
the Coulomb gauge magnetic helicity density,
say $\tilde{h} = \overline{\aaaa\cdot\bb}$,
would have $(k_k + K_k/2)/(\kk + \KK/2)^2$ replacing $k_k/k^2$
in the Fourier space expression of \Eq{helfour}. The two expressions are
identical for the homogeneous case, and even in the weakly inhomogeneous case
up to first order terms in $K/k$, but not in general. 
So, $h(\RR) \neq \tilde{h}(\RR)$ in general.

In order to compute the magnetic helicity 
evolution, we use the induction equation for $\bb$ in Fourier space,
$\partial\hat{b}_i(\kk)/\partial t
=-\epsilon_{ipq} \ii k_p \hat{e}_q$. Here 
$\hat{e}_q$ is the Fourier transform of 
the small scale electric field $e_q$, which is given by
(e.g., Moffatt 1978)
\EQ
\ee = -\uu \times \meanBB -\meanUU \times \bb
-\uu\times\bb + \overline{\uu\times\bb} + \eta\jj.
\label{ee}
\EN
Substituting this in the time derivative
of \Eq{helfour}, we get after some straightforward algebra,
\EQA
\frac{\partial h(\RR)}{\partial t} &=& \int \int \Big\{ 
-2\int \overline{\hat{e}_q(\kk + \half\KK) \hat{b}_q(-\kk +\half\KK)}
\nonumber \\
&+& 2 (K_jk_q/k^2) \ \overline{\hat{e}_q(\kk + \half\KK) \hat{b}_j(-\kk +\half\KK)}
\nonumber \\
&-& (K_sk_s/k^2) \ \overline{\hat{e}_q(\kk + \half\KK) \hat{b}_q(-\kk +\half\KK)}
\Big\} \nonumber \\
&\times&  {\rm e}^{i\KK\cdot\RR} \ \dd^3K \ \dd^3k .
\label{hevol}
\ENA
We denote the integrals over the three terms in curly brackets above as
$A_1$, $A_2$, and $A_3$, respectively.
From the definition of $\Phi$ in Eq.~(\ref{phi_def}), the first term is simply
$A_1 = -2\overline{\ee\cdot\bb}$, or
\EQ
A_1 = 2\overline{\bb\cdot(\uu \times \meanBB)}  
- 2\eta\overline{\jj\cdot\bb}
=-2\meanemf\cdot\meanBB-2\eta\overline{\jj\cdot\bb},
\EN
where $\meanemf=\overline{\uu\times\bb}$ is the turbulent EMF.
Note that for homogeneous turbulence only the term $A_1$ survives. This
is because the other two terms which involve a large $K_i$ in the integrand,
will introduce a large scale $R_i$ derivative when evaluating
the integral over $\KK$, and this vanishes in the homogeneous case.
So, for the homogeneous case we recover a local
generalization (without volume integration) of the  
magnetic helicity conservation equation (Brandenburg \& Subramanian 2005a)
\EQ
\partial h/\partial t = -2\meanemf\cdot\meanBB-2\eta\overline{\jj\cdot\bb}
\quad\mbox{(from $A_1$ term only).}
\label{ClosedSystem}
\EN
In the inhomogeneous case, since the terms $A_2$ and $A_3$ are scalars
that depend on $K_i$ in the integrand,
and hence a large scale $R_i$ derivative, they will
contribute purely to the flux of helicity. The only term that involves
volume generation of the helicity density is the $A_1$ term, which we see
involves correlations no higher than the two-point one.
This is in contrast to the current helicity evolution, which
involved undetermined triple correlations in their volume generation;
see SB04.

Let us now evaluate the helicity fluxes given by $A_2$ and $A_3$.
This involves straightforward but tedious algebra. We
also work out the flux to the lowest order in the $R$ derivative.
There are again three main types of contributions due to different
parts of the electric field $\ee$. First there is a contribution
proportional to $\meanBB$ due to that part $\ee = -\uu \times \meanBB +...$\ .
In Fourier space this gives $\hat{e}_q(\kk) = \epsilon_{qlm} \int 
\hat{u}_m(\kk - \kk')\hat{\meanB}_l(\kk') \dd^3k'$. We substitute
this into the expressions for $A_2$ and $A_3$, change the variables
to $\KK' = \KK - \kk'$, use the definition for $\chi_{ij}$ and evaluate
the integrations over $\KK'$ and $\kk'$ retaining only terms
to lowest order in the $R$ derivatives. We then get
$A_2= -\nabla_j \meanF^{\rm VC}_j$ and $A_3= -\nabla_j \meanF^{\rm A}_j$,
where the mean field dependent fluxes
$\meanF^{\rm VC}_j$ and $\meanF^{\rm A}_j$ are given by
\EQA
\meanF_i^{\rm VC} &=& 2\epsilon_{qlm} \meanB_l(\RR) \int
\ii k_q \chi_{mi} k^{-2} \dd^3k , \nonumber \\
\meanF_i^{\rm A} &=&  - \epsilon_{qlm} \meanB_l(\RR) \int
\ii k_i \chi_{mq} k^{-2} \dd^3k .
\label{fluxB}
\ENA
Note that $\meanF_i^{\rm A}$ only depends on the
antisymmetric part of the cross correlation $\chi_{mq}$,
whereas $\meanF_i^{\rm VC}$ is sensitive to the symmetric part as well.
Now consider the contribution proportional to the mean velocity
from the part of the electric field $\ee = 
 -\meanUU \times \bb + ...$\ . The evaluation of this follows the same
steps as in evaluating \eq{fluxB} except that one can map
$u_m \to b_l$ and $\meanB_l \to \meanU_m$. This gives
$A_2+A_3= -\nabla_j \meanF^{\rm bulk}_j$, where the flux due to bulk motions
is given by 
\EQ
\meanF_i^{\rm bulk} = \epsilon_{qlm} \meanU_m(\RR) \int \left(
2\ii k_q m_{li} -
\ii k_i m_{lq} \right) k^{-2} \dd^3k .
\label{fluxU}
\EN
Indeed, if the magnetic correlations were isotropic
then it is easy to simplify this further and one gets
$\meanF_i^{\rm bulk}\!=h\meanU_i$, exactly as one should for an advective flux!

The contribution to the fluxes from $\ee = 
-\uu\times\bb$ (see \Eq{ee}, introduces triple correlations in
the flux, which then need a closure theory to evaluate.
We denote this flux term as $\meanF^{\rm triple}_i$.
However, since this triple correlation comes only in the flux,
and not the volume generation term, it is likely that its
value can be constrained by a conservation law.
This will be examined in more detail in the future.
Note also that the contribution to the helicity flux from
$\ee = \overline{\uu\times\bb} + ..$ is zero and that
the resistive contribution from $\ee = \eta\jj + ...$ is likely
to be negligible compared to the terms that we retain.

Putting all our results together we can write for the evolution
of the magnetic helicity density
\EQ
\partial h/\partial t + \nab\cdot\meanFF
= -2\meanemf\cdot\meanBB-2\eta\overline{\jj\cdot\bb},
\label{finhel}
\EN
where the flux is
$\meanF_i=\meanF^{\rm VC}_i+\meanF^{\rm A}_i+\meanF^{\rm bulk}_i+\meanF_i^{\rm triple}$.
We should emphasize that \Eq{finhel} is a {\it local} magnetic 
helicity conservation law. If one were unable to define a gauge-invariant 
magnetic helicity density, one would
only have an integral (global) conservation law, as in previous studies.
Further simplification of the helicity fluxes 
for use in say MFD models, requires the evaluation
of the turbulent EMF tensor $\chi_{ij}$, which can be
done only under a closure scheme, 
and will be presented elsewhere. It turns out that 
$\meanF^{\rm VC}_i$ is a generalization of the magnetic helicity
flux obtained by Vishniac \& Cho (2001),
which is particularly important in the presence of strong shear 
(Brandenburg \& Subramanian 2005b).
The fluxes used by Kleeorin et al.\ (2000) can arise from
both $\meanF^{\rm A}_i$ and the contributions of the antisymmetric parts of the
correlations to  $\meanF^{\rm VC}_i$.
The magnetic helicity fluxes also vanish if the
turbulence is homogeneous. For isotropic but inhomogeneous turbulence,
$\meanF_i^{\rm VC}$ also depends purely on the antisymmetric
part of $\chi_{ij}$, just like $\meanF_i^{\rm A}$.
Furthermore, the VC and $A$-fluxes are proportional to two-point 
correlations ($\chi_{ij}$)
and the mean field $\meanBB$, see \Eq{fluxB};
this is unlike the two-dimensional case (Silvers 2006).

Since the small scale magnetic helicity opposes the
kinetic part of the $\alpha$ effect (Pouquet et al.\ 1976),
its loss through corresponding magnetic helicity fluxes
can alleviate this quenching effect 
(Blackman \& Field 2000; Kleeorin et al.\ 2000; Vishniac \& Cho 2001).
Note that the $\alpha$ effect quantifies the contribution of $\meanemf$
that is aligned with the mean field, i.e.\ $\meanemf=\alpha\meanBB+...$
for the simplest case of a scalar $\alpha$ effect.
For a closed system \Eq{ClosedSystem} applies and,
in the stationary limit, this predicts $\meanemf\cdot\meanBB
= -\eta\overline{\jj\cdot\bb}$ which tends to zero as $\eta \to 0$
for any reasonable spectrum of current helicity. This leads to a catastrophic
quenching of the turbulent EMF parallel to $\meanBB$.
In the presence of helicity fluxes, however, we have 
$\meanemf\cdot\meanBB = -\half\nab\cdot\meanFF
 -\eta\overline{\jj\cdot\bb}$ in the stationary limit,
and so $\meanemf\cdot\meanBB$ need not be catastrophically quenched.
The turbulent magnetic helicity fluxes worked out here
are therefore crucial for the efficient working of the mean field dynamo.
Numerical work in determining the $\alpha$ effect did show a 30-fold
increase in simulations that allowed helicity fluxes
to develop (Brandenburg \& Sandin 2004).
However, a more convincing demonstration of the importance of helicity
fluxes comes from a dynamo simulation in the presence of shear showing
that only with open boundaries can a significant large scale field of
equipartition field strength develop (Brandenburg 2005).

\section{Conclusions}

We have proposed here a local gauge invariant definition of
magnetic helicity density for random fields in weakly inhomogeneous systems,
which can also have boundaries. This is particularly useful in the context of
MFDs since one can then meaningfully discuss magnetic helicity fluxes
and the local effect of Lorentz forces.
We have derived an evolution equation for the
local magnetic helicity density and showed
that they naturally involve helicity fluxes, which may alleviate
the problems associated with MFDs (Shukurov et al.\ 2006).
Our work therefore lays the 
conceptual foundation for the many discussions of the effects of helicity
fluxes already existing in the literature and for future explorations.

Future applications might include the use of \Eq{heldef} together
with an assumption of isotropy to estimate the spatial variation of
magnetic helicity density by measuring at least one of the off-diagonal
components of $M_{ij}(\rrr,\RR)$.
This type of approach has been adopted by Matthaeus et al.\ (1982) to
determine the magnetic helicity in the solar wind by measuring just a
time series of an off-diagonal component of $M_{ij}(\rrr)$ under the
Taylor hypothesis.
However, no dependence on the large scale coordinate $\RR$ has been
determined.
In principle, similar ideas could be applied to determine $h(\RR)$ on
the solar surface without necessarily having access to the dependence
of the magnetic field with depth.

\begin{acknowledgments}
We thank the organizers and participants
of the program ``Magnetohydrodynamics of Stellar
Interiors'' at the Isaac Newton Institute in Cambridge (UK) for
a stimulating environment that led to the present work.
KS thanks NORDITA for their hospitality during the course of this work.
\end{acknowledgments}

%\vfill\bigskip\noindent{\it
%$ $Id: paper.tex,v 1.105 2006/08/30 06:54:45 brandenb Exp $ $}

\end{document}